\documentclass[Afour,sagev,times]{sagej}

\usepackage{moreverb,graphicx,placeins}

\usepackage[hyphens]{url}

\newcommand\BibTeX{{\rmfamily B\kern-.05em \textsc{i\kern-.025em b}\kern-.08em
T\kern-.1667em\lower.7ex\hbox{E}\kern-.125emX}}

\def\volumeyear{2015}

\begin{document}

\runninghead{Dykes et al.}

\title{Splotch: porting and optimizing for the Xeon Phi}

%\itshape{SAGE Publications}

\author{Timothy Dykes\affilnum{1}, Claudio Gheller\affilnum{2}, Marzia Rivi\affilnum{3}\affilnum{4} and Mel Krokos\affilnum{1}}

\affiliation{\affilnum{1}School of Creative Technologies, University of Portsmouth, Portsmouth, UK\\
\affilnum{2}CSCS-ETHZ, Lugano, Switzerland\\
\affilnum{3}Department of Physics and Astronomy, UCL, London, UK\\
\affilnum{4}Department of Physics, University of Oxford, Oxford, UK}

%\author{Anonymized \affilnum{1}, Anonymized\affilnum{2}, Anonymized\affilnum{3}\affilnum{4} and Anonymized\affilnum{1}}

%\affiliation{\affilnum{1}Anonymized\\
%\affilnum{2}Anonymized\\
%\affilnum{3}Anonymized\\
%\affilnum{4}Anonymized}
%\corrauth{Anonymized}
%\email{Anonymized}

\corrauth{Timothy Dykes, 
School of Creative Technologies,
University of Portsmouth,
Elton Building, Winston Churchill Avenue,
Portsmouth,
PO1~2DJ, UK.}

\email{timothy.dykes@port.ac.uk}

\begin{abstract}
With the increasing size and complexity of data produced by large scale numerical simulations, it is of primary importance for scientists to be able to exploit all available hardware in heterogenous High Performance Computing environments for increased throughput and efficiency. We focus on the porting and optimization of Splotch, a scalable visualization algorithm, to utilize the Xeon Phi, Intel's coprocessor based upon the new Many Integrated Core architecture. We discuss steps taken to offload data to the coprocessor and algorithmic modifications to aid faster processing on the many-core architecture and make use of the uniquely wide vector capabilities of the device, with accompanying performance results using multiple Xeon Phi. Finally performance is compared against results achieved with the GPU implementation of Splotch.
\end{abstract}


\maketitle

\section{Introduction}
\label{sect:introduction}
Nowadays dealing with large data effectively is a mandatory activity for a rapidly increasing number of scientific communities, e.g. in environmental, life and health sciences, and in particular in astrophysics. Some of the largest cosmological N-body simulations can describe the evolution of our universe up to present time by following the behavior of gravitating matter represented by many billions of particles. Performing such simulations often produces data outputs (or ``time snapshots'') in the order of multiple terabytes. This situation can only be exacerbated as advances in supercomputing are opening possibilities for simulations producing snapshots of sizes in the order of petabytes, or even exabytes to look toward the exa-scale era.

Large size is not the only challenge posed, it is also essential to effectively extract information from the typically complex datasets. Algorithms for data mining and analysis are often highly computationally demanding; visualization can be an outstanding analytical aid for further exploration and discovery, e.g. by providing scientists with prompt and intuitive insights enabling them to identify relevant characteristics and thus define regions of interest within which to apply further intensive and time-consuming methods, thereby minimizing unnecessary expensive analysis. Furthermore, it is a very effective way of qualitatively discovering and understanding correlations, associations and data patterns, or in identifying unexpected behaviors or even errors. However, visualization algorithms typically require High Performance Computing (HPC) resources to overcome issues related to rendering large and complex datasets in acceptable timeframes.

Splotch~\cite{Splotch01} is an algorithm for visualizing large particle-based datasets, providing high quality imagery while exploiting a broad variety of HPC systems such as multi-core processors, multi-node supercomputing systems~\cite{Splotch02}, and also GPUs~\cite{Splotch03}. The variety of implementations is due to the fact that many HPC systems of today are exploiting not only standard CPUs, but accelerators to achieve maximum computational power with low energy usage. As of November 2014, five of the top ten supercomputers in the world exploit GPUs or Xeon Phi coprocessors, including the number one supercomputer Tianhe-2 (developed by China's National University of Defense Technology) which runs on a combination of 12-core Intel Xeon E5 CPUs and Xeon Phi 31S1P devices\cite{Top500}. The ability to fully exploit modern heterogenous HPC systems is of paramount importance towards achieving optimal overall performance. To this end, this paper reports on recent developments enabling Splotch to exploit the capability of the Intel Xeon Phi~\cite{XeonPhi01} coprocessor, taking advantage of the Many Integrated Core (MIC) architecture~\cite{XeonPhi02}, which is envisaged to provide, on suitable classes of algorithms, outstanding performance with power consumption being comparable to standard CPUs.

Many developers have been exploring the possibility of using the MIC based products to accelerate their software, as can be seen in the Intel Xeon Phi Applications and Solutions catalogue~\cite{XeonPhi03}; high performance visualization software developers are also on board with efforts being made to extend VisIt\footnote{\url{https://visit.llnl.gov}}, an open source scientific visualization tool, to exploit the MIC architecture at the Intel Parallel Computing Center at the Joint Institute for Computational Sciences between the University of Tennessee and Oak Ridge National Laboratory. Developers are enticed not only by the large potential for compute power, but also a key advertised feature of the Xeon Phi, and MIC architecture: the ability for developers to work with regular programming tools and methods they would use for a standard Xeon (or indeed, other processors). This includes use of parallel APIs and runtimes such as MPI and OpenMP from standard C++ and Fortran  source code, while also offering the possibility of using more specialized options for parallelism such as Intel's Cilk Plus\footnote{\url{http://www.cilkplus.org/}}.

Current experiences implementing algorithms or porting pre-existing parallel codes to the Xeon Phi show many successes optimizing specific kernels and improving performance as compared to unoptimized kernels on Xeon Phi (e.g.  \cite{GadgetMicPrace}, \cite{XeonPhiCFD}). Despite this, when comparing overall program performance against that achievable on a node with standard Xeon CPUs, authors may achieve similar or lower performance to the CPU (e.g.~\cite{SmeagolXeonPhi}, \cite{CP2KMicPrace}), and it is not surprising considering the recent introduction of the architecture and the reported difficulties in achieving performance \cite{TestDrivingXeonPhi}. This, however, is in the process of changing as more efforts are made to utilize the architecture with improvements showing in all areas as the technology matures. One particular success story of note is the exploitation of the Xeon Phi enabled Tianhe-2 to achieve peta-scale performance with an earthquake modeling simulation~\cite{PetaScaleMic}. It is expected that further success stories will appear as more machines are built exploiting this architecture and more developers investigate this area of technology, especially considering the commitment made to the architecture with new supercomputers featuring Xeon Phi \emph{Knights Landing} chips commissioned by both the US Department of Energy's NERSC Centre and the NNSA's Advanced Simulation and Computing (ASC) program, with delivery expected in 2016. 

In view of this, throughout this paper we explore the suitability of the Splotch visualization algorithm for the Xeon Phi, in order to be ready to take advantage of large many-core systems. We share our experiences implementing an offloaded processing model, and applying optimizations as outlined in the various Intel guides available, along with ideas of our own in regards to memory management and vectorization. With the rapidly increasing number of cores available in Xeon processors, many of these optimizations can also be applied on a Xeon CPU, and so we can also look to improve the OpenMP and MPI implementations of Splotch based on lessons learned throughout this experience.

The structure of the paper is as follows: we provide a brief background to Splotch and the Xeon Phi (Sect.~\ref{sect:background}), describe our MIC implementation (Sect.~\ref{sect:micsplotch}) focusing on optimization issues related to memory usage, data transfers, and vectorization along with discussion of performance analysis methods with some examples. We then discuss the performance details (Sect.~\ref{sect:results}) of our implementation using a benchmark dataset produced by a Gadget\footnote{The Gadget code: \url{http://www.mpa-garching.mpg.de/gadget/}} N-Body simulation and a cluster of up to sixteen Xeon Phi devices. We include a comparison of performances achieved implementing the Splotch algorithm separately for Xeon Phi and GPU (Sect.~\ref{sect:micvsgpu}), and finally in Sect.~\ref{sect:conclusions} we summarize our experiences and present pointers to future developments.

\section{Background}
\label{sect:background}

\subsection{The Splotch Code}
\label{sect:splotchcode}

\begin{figure} 
\centering
 \fbox{
  \includegraphics[height=4cm]{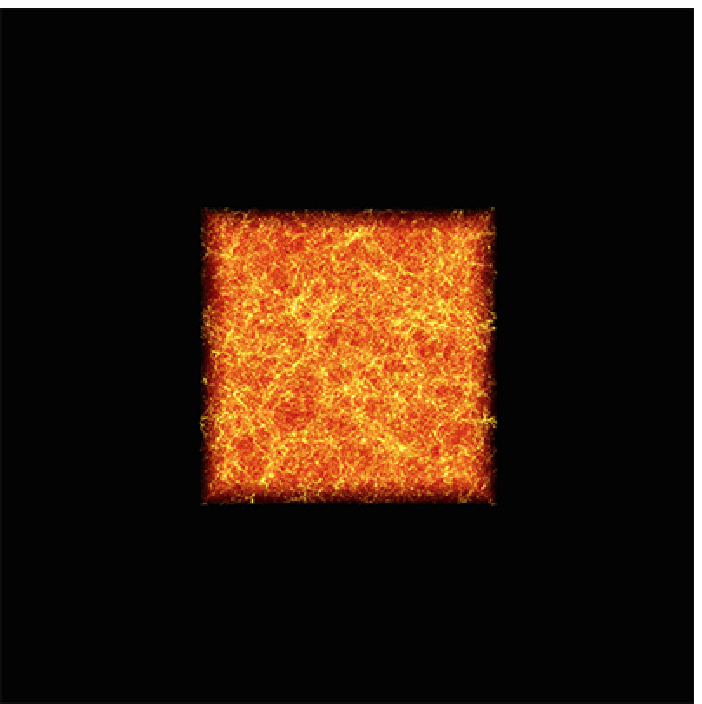}
  \includegraphics[height=4cm]{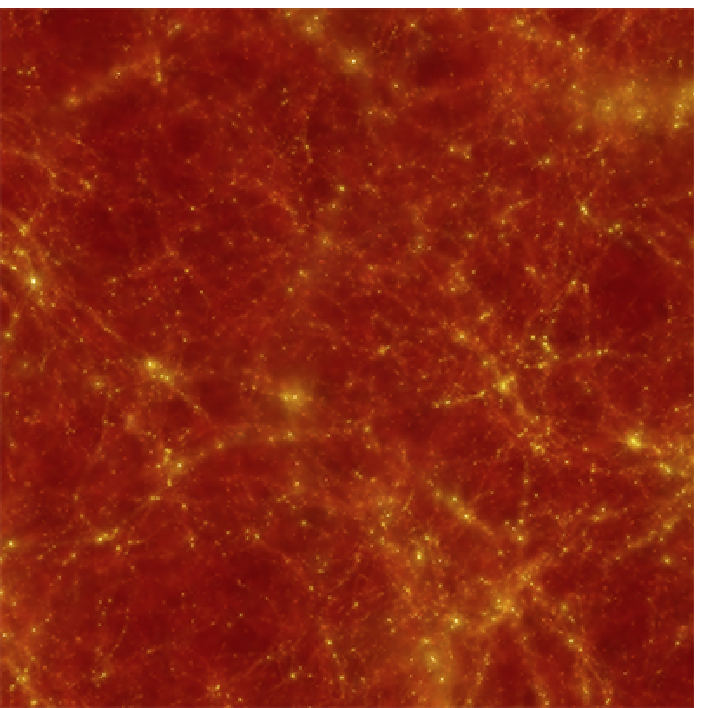}
  }
\caption{Two Splotch images of the dataset used for performance testing, seen from the closest distance (\emph{right}) and most remote distance (\emph{left}). The images depict a data cube describing large scale universe structure produced by an N-Body SPH simulation using the Gadget code. }
\label{fig:visualization}
\end{figure}

\begin{figure*} 
\centering
\fbox{
\includegraphics[height=8.0cm]{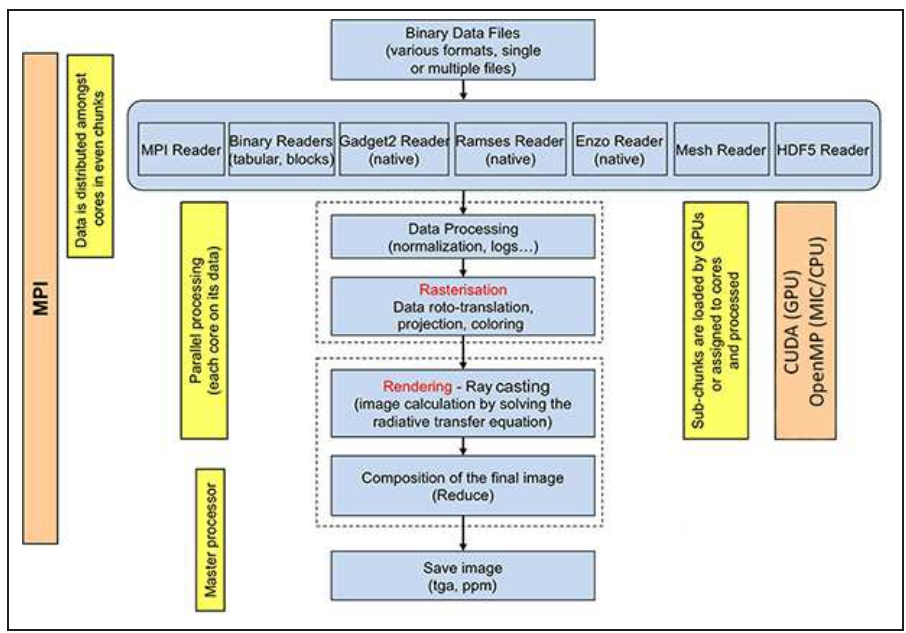}
}
\caption{Standard execution model of the Splotch code.}
\label{fig:splotchmodel}
\end{figure*}

Splotch \footnote{\url{https://github.com/splotchviz/splotch/wiki}} is implemented in pure C++, with no dependencies upon external libraries except where necessary for external file formats (e.g. HDF5\footnote{\url{http://www.hdfgroup.org/HDF5/}}), and includes several readers supporting a number of popular formats for astrophysics. Figures~\ref{fig:visualization} and~\ref{fig:splotchmodel} illustrate the type of visualization produced by Splotch and the current execution model of the Splotch algorithm; datasets are converted into the Splotch internal format as they are loaded from files, followed by preprocessing, rasterization, and finally rendering.

Preprocessing performs ranging, normalization, and optionally apply common functions to input fields (e.g. logarithm). At the rasterization stage, particle positions represented as 3 dimensional cartesian coordinates are transformed (roto-translated), and a perspective projection is applied with reference to supplied camera and look-at positions, followed by clipping. Colors, stored as a triplet of single precision floating point values representing red, green, and blue components (RGB) are generated either directly from a vector input field or as a function of a scalar input field using a color map loaded at runtime. For rendering, contrary to traditional rasterization scan conversion methods, rays are cast along lines of sight and contributions of all encountered particles are calculated based on the radiative transfer equation (for a more detailed description see~\cite{Splotch01}). 
%Perhaps some note on similarity to 'splatting'?

The area of the image influenced by a particle is dependent on the radius, an intrinsic scalar component of each particle, and the distribution of the affected pixels is obtained through a gaussian kernel. A larger average radius throughout the dataset results in longer rendering times, as this indicates particles affect larger portions of the image, this is caused for example by moving the camera close to or within the dataset. 

One notable feature of Splotch is the ability to supply a \emph{scene file}, which provides the ability to modify parameters, e.g. the brightness or color palette thresholds, over a progression of images which can then be stitched together to form a movie, allowing the user to highlight evolutionary properties of the simulated dataset. Figure~\ref{fig:visualization} shows some of the Splotch visualizations of the sample dataset used for performance analysis.

\subsection{Overview of the Xeon Phi}
\label{sect:micoverview}

The idea behind MIC is obtaining a massive level of parallelism for increasing throughput performance in power restricted cluster environments. To this end Intel's flagship MIC product line, the Xeon Phi, contains roughly 60 cores on a single chip, dependent on the model, and acts as an accelerator for a standard Intel Xeon processor. Programs can be executed natively by logging via ssh into the device, which hosts a Linux micro-OS, or by using the device through one or more MPI processes in tandem with those running on the Xeon host (symmetric mode). Alternatively users can offload data and portions of code to the coprocessor via Intel's Language Extensions for Offload (LEO), a series of pragma based extensions available in C++ or Fortran. For a detailed technical description of the processor's architecture, the reader is referred to the Xeon Phi white paper~\cite{XeonPhi01}. Here, we give a very short overview of its main features.

Each core has access to a 512 KB private fully coherent L2 cache and memory controllers and the PCIe client logic can access up to 8 GB of GDDR5 memory. A bi-directional ring interconnect brings these components together. The cores are in-order and up to 4 hardware threads are supported to mitigate the latencies inherent with in-order execution. The Vector Processor Unit (VPU) is worth mentioning due to the innovative 512 bit wide Single Instruction Multiple Data (SIMD) capability, allowing 16 single precision (S.P.) or 8 double precision (D.P.) floating point operations per cycle, with support for fused-multiply-add operations increasing this to 32 S.P. or 16 D.P. floating point operations.

\FloatBarrier
\section{Splotch on the MIC}
\label{sect:micsplotch}

\subsection{Implementation}
\label{sect:micimplementation}

\begin{figure}
\centering
\fbox{
\includegraphics[height=7.0cm]{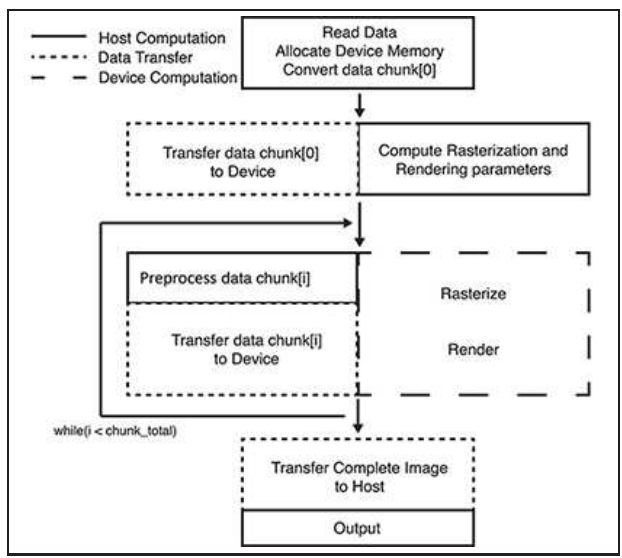}
}
\caption{Model illustrating execution flow of the offloading implementation.}
\label{fig:exmodel}
\end{figure}

The Splotch algorithm ported to the Xeon Phi uses the offload model, illustrated in Fig.~\ref{fig:exmodel}. Whilst the executable runs on the Xeon host, data and processing are offloaded to the device via Intel's LEO. The advertised ability to run already OpenMP and MPI based programs on the Xeon Phi has led to the logical decision to modify the existing code to run effectively on the device, as opposed to moving to another software paradigm such as Intel's Cilk Plus which may provide additional features but may involve a more in-depth modification of the algorithm, future work is envisioned to also explore this paradigm for further comparison.

A problem that is often apparent when exploiting coprocessors is the limited device memory. To facilitate the visualization of datasets potentially much larger than the memory capacity available, while minimizing overhead due to data transfers, a double buffered scheme has been implemented using the ability to asynchronously transfer data via a series of \emph{signal} and \emph{wait} clauses provided as part of the Intel C++ LEO extensions. This scheme creates two storage buffers; data from the host is copied to the first buffer, and while this data is processed the second buffer is asynchronously populated. The second buffer is then processed while the first buffer is asynchronously replaced with a new set of data. This loop can continue indefinitely while data is available in host memory, thereby solving the problem of limited device memory without costly delays in processing due to waiting for a full device buffer to be repopulated.

The rasterization phase consists of a highly parallel 3D transform, projection, clipping, and color assignment on a per-particle basis. Transform parameters are precomputed asynchronously, and work is distributed amongst threads via two OpenMP parallel for-loops, firstly for the transform and secondly for the colorize. The already highly parallel nature of these loops meant that no significant algorithmic modifications were needed in order for the code to run, however both are optimized through use of manual and automatic vectorization to provide a performance boost for this phase (see Sect.~\ref{sect:vectorization}).

For the rendering phase, the image is split into a two dimensional grid of tiles which are distributed amongst threads. The size of these square tiles are defined by a parameter provided at runtime (see Sect.~\ref{sect:tuning} for more). A list of particle indices is created for each tile, indicating all particles affecting that tile. Each thread renders a full list by solving the radiative transfer equation, and retrieves another in round-robin fashion; in this way pixel access is kept local to each tile and not shared between threads, avoiding concurrent access and race conditions. Finally when all chunks of data have been processed and accumulated, the resultant device image is copied back to the host for output. The original OpenMP rendering process described in~\cite{Splotch03} has not been conceptually modified, rather the implementation has been optimized for Many Integrated Core. All further references to rendering (e.g. for performance testing) refer to the full process of particle-to-tile distribution and radiative transfer computation.

\subsection{Optimization}
\label{sect:micoptimization}

% Perhaps here some introduction to how the optimization is approached, i.e. an architectural optimization
A set of key problems known to cause performance issues with the Xeon Phi were targeted, as laid out in various resources for programming such devices \footnote{For example: \url{https://software.intel.com/en-us/mic-developer}}. Memory usage, data transfer, vectorization, and general tuning are discussed in the following subsections. In addition we used two key tools to analyze performance and identify target areas for optimization. 

The first is Intel VTune Amplifier XE, a profiling tool capable of measuring, analyzing, and visualizing performance of processing, both offloaded and native, on the Xeon Phi. 
%A useful aspect of this tool is the ability to display the length of time spent executing each statement, while also comparing the statement with corresponding assembly code. This facilitates targeted optimization of small hotspots, allowing to quickly see if a particular statement is problematic. 
Amongst other features, it simplifies the process of evaluating the benefit of manually inserted intrinsics in comparison to those generated by the compiler. 
%Metrics, derived or core, can be taken fairly easily and there is the possibility to auto-detect issues and hint at possible solutions. 

The second tool we adopted for tuning is the Performance API (PAPI)\cite{PAPI}. This API assists direct measurement of hardware events on the device. In particular the ability to target a selected set of statements with in-code hooks can be used for fine grained capture of hardware events in a large codebase. Moreover PAPI provides a high level of hardware event control in a lightweight form, without the auto-analysis and visualization options of a more fully featured tool such as VTune Amplifier. We used a small wrapper to facilitate use of PAPI in Intel's offload mode, which is available on Github with a sample benchmark and setup instructions\footnote{\url{https://github.com/TimDykes/mic_papi_wrapper}}. 
%\footnote{Anonymized}.

\subsubsection{Memory Usage and Data Transfer}
\label{sect:memusage}

Cost of dynamic memory allocation on the Phi is relatively high~\cite{IntelDocs01}, so in order to minimize unnecessary allocations buffers are created at the beginning of the program cycle and reused throughout. We found that we were unable to asynchronously allocate memory using offload clauses (allocating directly with a LEO clause as opposed to offloading a call to \emph{malloc}), and so overheads incurred allocating these buffers cannot be mitigated by overlapping allocation with host activity.
Use of the MIC\_USE\_2MB\_BUFFERS environment variable forces buffers over a particular size to be allocated with 2MB pages rather than the default 4KB, which improves data allocation and transfer rates and can benefit performance by potentially reducing page faults and translation look-aside buffer misses~\cite{IntelDocs02}. We have performed tests using offload clauses for compiler managed memory allocation on the device. To our experience a single process offloading to the device, and reserving large buffers, can allocate memory roughly 2-2.5x faster having set this environment variable appropriately. 

\begin{figure*}
\centering

\fbox{
\includegraphics[height=6.0cm]{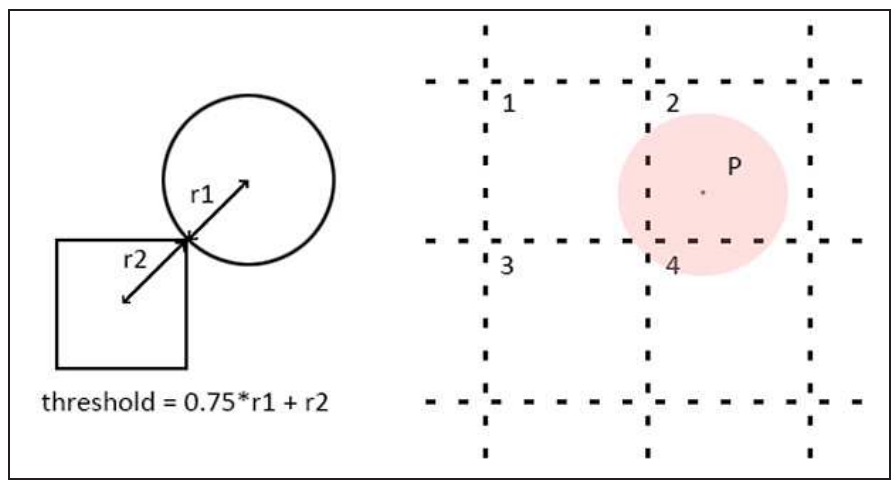}
}
\caption{\emph{Left:} Distance threshold definition below which a particle is considered to affect a tile. 
\emph{Right:} Example of a particle projected onto a 2D tiled image. In this case, the particle is considered to affect tiles 1, 2, and 4, as the overlap between particle~P and tile~3 is negligible.}
\label{fig:tilesdiagram}
\end{figure*}

A notable issue for the many-core architecture is scalable memory allocation. When working with dynamic memory, for example through use of Standard Template Library (STL) containers, parallel calls to \emph{malloc} can be serialized~\cite{IntelDocs03}.  In our case, this can occur at the beginning of the rendering phase, where the target image is geometrically decomposed into $n$ tiles and for each tile a vector of particle indices is generated (See Fig's  \ref{fig:tilesdiagram} and \ref{fig:pseudocode}). This requires each OpenMP thread to create $n$ arrays, which are then filled with particle indices to be rendered. Each vector requests more memory as it reaches capacity, resulting in significant stalls caused by simultaneous allocations from different threads,.

\begin{figure}
\centering
\small
\newcommand{\keyw}[1]{{\bf #1}}
\begin{tabbing}
\quad \=\quad \=\quad \=\quad \=\quad \=\ quad \kill

\keyw{begin parallel block} \\
\> get\_particles\_for\_thread$(threadID, start, end)$ \\

\>\keyw{for each} $p$ \keyw{in} $particles[start]$ : $particles[end]$ \\
\>\> get\_affected\_tiles$(range, tile\_size, p.x, p.y, p.radius)$ \\
\> \>$threshold$ = 0.75*$p.radius$ + sqrt(2)*$tile\_size$/2	 \\
\> \> \keyw{for each} $x$ \keyw{in} $tiles[range.min\_x]$ : $tiles[range.max\_x]$ \\
\> \> \>$cx$ = get\_center$(x)$ \\
\> \> \>\keyw{for each} $y$ \keyw{in} $tiles[range.min\_y]$ : $tiles[range.max\_y]$ \\
\> \> \> \>$cy$ = get\_center($y$) \\
\> \> \> \>$distance$ = length$((p.x, p.y),(cx, cy))$ \\
\> \> \> \>\keyw{if}($distance$~\textless~$threshold$) \\
\> \> \> \> \>$indices\_vector[thread\_id][x][y]$.push\_back$(p.index)$ \\
\> \> \> \>\keyw{end if}\\
\> \> \>\keyw{end for}\\
\> \>\keyw{end for}\\
\>\keyw{end for}\\
\keyw{end parallel block}
\end{tabbing}
\caption{ Pseudocode illustrating the construction of vectors of particle indices which affect the tiles of a decomposed image. }
\label{fig:pseudocode}
\end{figure}

%In our case roughly 625 tiles $\times$ 236 threads, making almost 150000 arrays compared to roughly 2000 on a Xeon which uses lower numbers of threads and tiles. Therefore the likelihood of stalled calls to \emph{malloc} is higher, and this coupled with slower allocation on the Xeon Phi can cause a significant performance decrease as illustrated in Fig.~\ref{fig:allocator}. 

In the original implementation no attention is paid to potential memory allocation stalls at this point due to the relatively low number of threads, typically 16 or less. The number of threads active on the many-core architecture however is often much higher, resulting in thousands of dynamic arrays across many threads, necessitating a scalable solution to memory allocation.
There are already solutions able to achieve sufficiently scalable memory allocation in parallel environments such as use of external libraries, e.g. \emph{ptmalloc}\cite{ptmalloc}, or constructs provided with parallel libraries, e.g. Intel Threading Building Blocks \emph{scalable\_allocator}\cite{tbballocator}. However, in order to solve this problem without introducing dependancies on external libraries, we implement a custom memory management solution consisting of a template array (\emph{array\_t}) and a memory allocator. Each thread is provided with an instance of the allocator, which asynchronously allocates a user-defined subset of storage from the device memory in the initial setup of the program. This ``pooled" memory is then sub-allocated on request to local arrays, which greatly reduces risk of clashing calls to \emph{malloc} from differing threads. The allocator implements bi-directional coalescence in order to minimise fragmentation, maintains the requested alignment for allocations, and if unable to provide the requested amount of memory it will request this via a standard aligned allocation. 

\begin{figure}
\centering
\fbox{
\includegraphics[height=5.0cm]{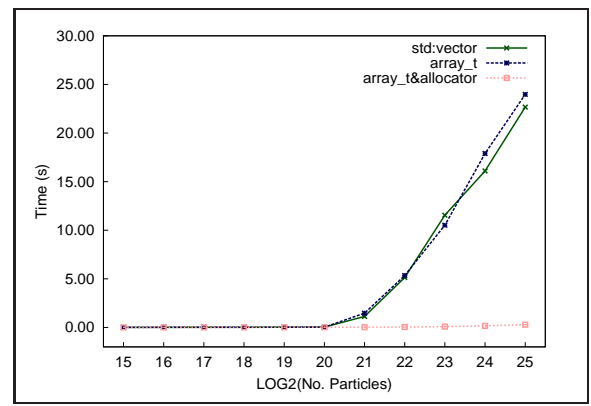}
}
\caption{Graph comparing performance of an \emph{std::vector} with a custom array implementation (\emph{array\_t}) with and without pooled memory allocation.}
\label{fig:allocator}
\end{figure}

We compare performance of an STL vector container with and without Intel's scalable\_allocator, against our custom vector implementation \emph{array\_t} with and without the pooled memory allocator. We can see in Fig.~\ref{fig:allocator} that performance for the STL container, and custom array implementation using standard aligned allocation methods (i.e. \emph{\_mm\_malloc()}), decreases significantly as the particle count rises above 2\textsuperscript{20}, or roughly 1 million. The \emph{array\&allocator} retains high performance due to all containers local to a thread being allocated memory from a dedicated per-thread memory pool. In the largest cases, where allocations begin to exceed local memory pools and are only limited by device memory, the custom allocator performs slightly better; this is in part due to faster 2MB page allocation, which does not appear to benefit Intel's scalable allocator.

% Does the intel one have this same property? per thread memory pool? probably
% Can we see memory usage??
% Check against cache aligned allocator (done: its not comparable)

    Overheads in dynamic allocation and data transfer can incur a penalty when running a single host process offloading to the device. In order to minimize these penalties, we make use of the MPI implementation of Splotch. Multiple MPI processes on the host are each allocated a subset of the device threads to exploit. In this way, the device is subdivided amongst the host MPI processes allowing to minimize overheads in data transfer, allocation and processing providing a noticeable performance increase, further details of which are given in Sect.~\ref{sect:results} and similar experiences can be seen in~\cite{GadgetMicPrace}. 
    
\subsubsection{Vectorization}
\label{sect:vectorization}

The large 512 bit width SIMD capability of the MIC architecture is exploited through vectorization carried out both automatically by the compiler, and manually using Intel intrinsics which map directly to Intel Initial Many-Core Instructions (IMCI)~\cite{IntelARM}.

The roto-translation and filtering stage of rasterization was modified to enable auto-vectorization through a combination of optimizations described in Intel's Vectorization guide~\cite{IntelVectorGuide}, and feedback from the compiler vectorization report.

 The coloring stage of rasterization involves a per-particle inner loop through an external color look-up table. This causes the compiler not to auto-vectorize the most suitable loop that in our case is the outer one through particles. As a solution, we iterate through the outer loop in increments of 16 (i.e. the device S.P. SIMD vector capability) and manually vectorize the iteration; the remainder loop (if any) is processed in a non-vectorized fashion. This improvement doubled the performance of this section of code as detailed below.  

\begin{table*} 
\small\sf\centering
\begin{tabular}{l r r r r}
\hline \noalign{\smallskip}
\textbf{Core Event} & \textbf{Before} & \textbf{After} & \textbf{\% Diff.} & \textbf{Effect}\\ \hline
\texttt{time} & 0.16 & 0.07 & -53.88 & positive\\ 
\texttt{vpu\_instructions\_executed} & 2.18E+09 & 1.49E+09 & -31.53 & positive\\ 
\texttt{vpu\_elements\_active} & 9.00E+09 & 6.60E+09 & -26.69 & positive\textsuperscript{a}\\
\texttt{cpu\_clk\_unhalted} & 3.54E+10 & 2.31E+10 & -34.72 & positive\\ 
\texttt{instructions\_executed} & 8.08E+09 & 6.24E+09 & -22.82 & positive\\ 
\texttt{long\_data\_page\_walk} & 8.94E+03 & 7.99E+03 & -10.61 & positive\\ 
\texttt{L2\_data\_read\_miss\_mem\_fill} & 6.69E+06 & 1.11E+07 & 66.51 & negative\\ 
\texttt{L2\_data\_write\_miss\_mem\_fill} & 1.15E+07 & 1.50E+05 & -98.69 & positive\\ \hline
\textbf{Derived Metric} &  &  &  & \\ \hline
\textit{Vector intensity} & 4.13 & 4.42 & 7.07 & positive\\ 
\textit{GFLOP/s} & 55.82 & 88.73 & 58.97& positive \\ 
\textit{Per-Thread CPI} & 4.38 & 3.7 & -15.42 & positive\\
\textit{Per-Core CPI} & 1.09 & 0.93 & -15.42 & positive\\ 
\textit{Read Bandwidth (bytes/clock)} & 0.13 & 0.13 & -4.73 & negative\\ 
\textit{Write Bandwidth (bytes/clock)} & 0.07 & 0.09 & 28.89 & positive\\ 
\textit{Bandwidth (GB/s)} & 12.57 & 13.46 & 7.07& positive \\ \hline
\end{tabular}\\
\textsuperscript{a}Positive when the percentage reduction in \texttt{vpu\_instructions\_executed} is larger.
\caption{Table comparing various hardware events and metrics measured with PAPI before and after manual vectorization of the colorize kernel. Both positive and negative values of the percentage differences can indicate a performance improvement, depending on the parameter. Changes which are beneficial for the algorithms are indicated as ``positive'' in the last column.}
\label{tab:papi_metric}
\end{table*}

As the rendering phase of the algorithm is unsuitable for automatic vectorization, this is also manually optimized through extensive use of intrinsics. Drawing consists of additively combining a pixel's current RGB values with the contribution from the current particle, which is calculated by multiplying the particle color by a scalar contribution value. In order to expedite this process, up to five single precision particle RGB values (totaling 480 bits) and five scalar contribution values are packed into two respective 512 bit vector registers. A third register contains 5 affected pixels, which are written simultaneously using a fused-multiply-add vector intrinsic, masked in order not to affect the final unused float value in the 16-float capable registers.

In all stages, PAPI was used to assist in ascertaining the effect of particular optimizations. An example can be seen in Table.~\ref{tab:papi_metric}; a set of native hardware events, along with derived metrics, have been measured before and after the insertion of manual intrinsics to optimize the colorize kernel. It can be seen from this example that the manual intrinsics doubled the speed of the kernel, while the measured FLOP/s increased by 60\%. 

In this case, likely contributors to the performance gain include the 20\% decrease in instructions retired per thread (event: \emph{instructions\_executed}) coupled with a 30\% drop in CPU cycles overall (event: \emph{cpu\_clk\_unhalted}). The 7\% rise in vector intensity (derived metric: \emph{vpu\_elements\_active} / \emph{vpu\_instructions\_executed}) indicates more of the vector elements in a VPU register were active on average during vectorized execution, leading to less instructions necessary. It is likely that the reformat of instructions involved in manual vectorization had a higher impact on the reduced cycle count (and therefore reduced time to solution) than the mildly higher level of vectorization indicated by the vector intensity metric. The authors note it can be difficult to interpret the meaning of hardware events when not accompanied by an increase or decrease in time, and so for more information on the capture, utilization and derivation of hardware metrics the reader is referred to the relevant Intel optimization guide~\cite{IntelDocs04}. 

The experience of trying to push the vectorization capabilities of the compiler and investigating different areas of the algorithm in an effort to optimize for vectorization has led to the authors recommendation that while the compiler can be very useful in automatically vectorizing code, it is still possible to gain significant performance boosts by manually inserting intrinsics to complex areas of code.

\subsubsection{Tuning}
\label{sect:tuning}

For relatively small datasets where processing time is low, i.e. a matter of seconds, initialization of the device and OpenMP threads can cause a noticeable overhead. The impact of this can be minimized by placing an empty offload clause with an empty OpenMP parallel section near to the beginning of the program, in order to overlap this overhead while other host activity is occurring, in this case while reading input data. Alternatively the environment variable OFFLOAD\_INIT can be set to pre-initialize all available MIC devices before the program begins execution. 

Various parameters of the algorithm can be tuned to find best performance. Render parameters such as the number of thread groups and tile size are set to optimal defaults for the test hardware based on results of scripted tests iterating through incremental sets of potential values. These can be modified via a parameter file passed in at runtime for differing hardware. 

\begin{figure}
\centering
\fbox{
\includegraphics[height=5.0cm]{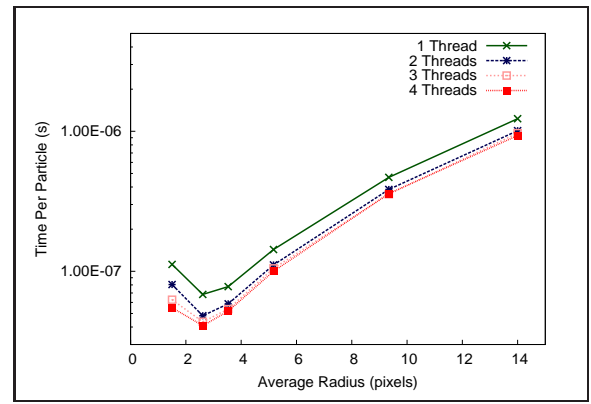}
}
\caption{Comparison of performance with one to four threads per core, using optimal \emph{scatter} affinity settings.}
\label{fig:threadCount}
\end{figure}

Thread count and affinity are important factors in the tuning process. We examine the effect of varying the number of threads per core and thread affinity for a series of Splotch renderings. 
We run a set of tests varying the camera position in order to have a fair comparison of the effect as a function of the average particle radius (see Fig.~\ref{fig:threadCount}). We tested with one to four threads per core, four being the maximum number of hardware contexts available per core. A series of preliminary tests indicated the \emph{scatter} affinity\footnote{For more on thread affinity see: \url{https://software.intel.com/en-us/articles/openmp-thread-affinity-control}} is ideal for our use case and so we set this configuration for the final tests,  
although we noted that in the case of 4 threads per core the difference between affinity settings (in particular \emph{scatter} and \emph{balanced}) was negligible most of the time, as also mentioned in~\cite{CP2KMicPrace}.

From Fig.~\ref{fig:threadCount} we see that the best performance is obtained for 4 threads per core. It is important to run these tests, which can be trivially scripted, for any multi-threaded application as it is likely the suitable configuration will be different from ours. It can also be noted that, as expected, the gap between one to two threads per core is noticeably larger than between two to three and three to four; this is likely due to the inability of the core to issue two instructions from a single hardware thread context in back to back cycles~\cite{IntelDocs05}.

\section{Results}
\label{sect:results}

\subsection{Hardware and Test Scenario}
\label{sect:testhardware}
All tests are performed using the Dommic\footnote{http://user.cscs.ch/computing\_resources/dommic/index.html} facility at the Swiss National Supercomputing Centre, Lugano. In this eight node cluster, each individual node is based upon a dual socket eight-core Intel Xeon 2670 processor architecture running at 2.6~GHz with 32~GB of main system memory. Two Xeon Phi 5110 MIC coprocessors are available per node, making up to sixteen \emph{Knights Corner} coprocessors available.

The sample dataset used for measurements is a snapshot of an N-Body SPH simulation performed using the Gadget code. For single node tests we filter to 50~million gas particles and 5~million star particles ($\sim$1.8~GB) in order to process a single chunk of data and more accurately measure individual kernels, whereas for multiple node and GPU comparison tests we use the full size of 200~million gas particles and 20~million star particles ($\sim$7.2~GB).

\subsection{MIC Performance}
\label{sect:micperf}
A 100~frame animation with the camera orbiting the data is used to measure average per-frame timings producing images of $1024^2$ pixels. For performance comparisons the device uses a tile size parameter of 40 pixels, the optimal value chosen via a series of scripted performance tests. The Xeon uses a larger tile size to more effectively utilize the lower number threads with the more powerful cores of the Xeon. 

For clarity, in all OpenMP tests on the Xeon we use one thread per core. For MPI offloading to device, each task is allocated an even share of the 236 hardware thread contexts on the device which are then exploited with OpenMP, and in this case thread binding is set explicitly. In the case of one task using the whole device, thread affinity is set as per the optimal settings for our use case (Sect.~\ref{sect:tuning}).
\FloatBarrier
\subsubsection{Single Node Speed-up}
\label{sect:singlenodeperf}

\begin{figure} 
\centering
\fbox{
\includegraphics[height=5.0cm]{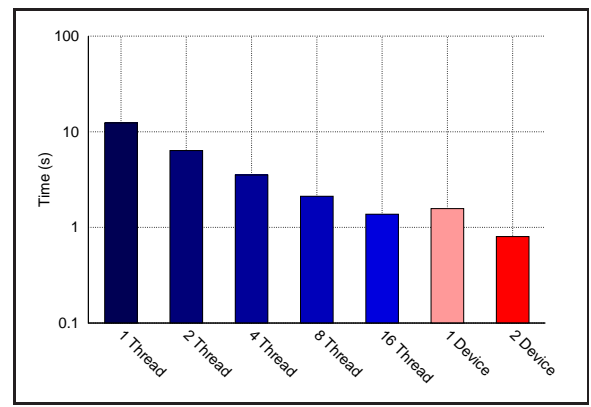}
}
\caption{Per-Frame total processing time: Xeon OpenMP 1-16 threads vs single and dual Xeon Phi devices.}
\label{fig:alltimings}
\end{figure}

\begin{figure} 
\centering
\fbox{
\includegraphics[height=5.0cm]{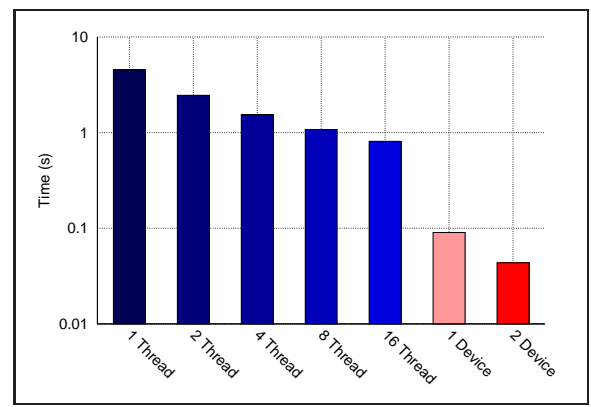}
}
\caption{Per-Frame rasterisation time: Xeon OpenMP 1-16 threads vs single and dual Xeon Phi devices.}
\label{fig:rotocoltimings}
\end{figure}

In order to accurately measure individual kernel performance we test with a relatively small dataset in a single node; in the test system this allows using two Xeon Phi devices. We compare performance of the OpenMP parallel version of Splotch on the Xeon and the Xeon Phi implementation exploiting both OpenMP and MPI.

\begin{figure} 
\centering
\fbox{
\includegraphics[height=5.0cm]{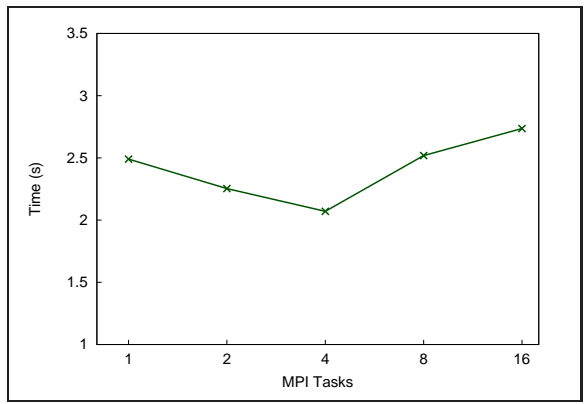}
}
\caption{Per-Frame processing time comparing multiple MPI tasks on the host sharing a single Xeon Phi; each task is further parallelised through OpenMP to use the full number of hardware thread contexts available.}
\label{fig:mpitimes}
\end{figure}

Figure~\ref{fig:alltimings}, describing per frame processing times of the OpenMP Xeon implementation vs dual and single devices, shows that use of a single device provides results close to 16 threads on the Xeon. Figure~\ref{fig:rotocoltimings} shows the strongest area of improvement, the rasterization phase, with a single device outperforming 16 threads (two CPUs) roughly 9x, with roughly 18x improvement provided by using dual devices. In both cases the use of a second device provides a 2x performance improvement for the MIC algorithm. The best performance in terms of FLOP/s is achieved in the transform kernel, which roto-translates, projects and filters particles as a subset of the rasterization phase. PAPI measurements indicate this kernel in a single device achieves $\sim$300~GFlop/s, or 15\% of peak S.P. performance. 

Figure~\ref{fig:mpitimes} shows comparison of per-frame processing times using varying numbers of MPI processes on the host, offloading OpenMP parallelised code to a single Xeon Phi. Subdividing the available device threads amongst MPI processes allows to more effectively spread the workload across the device to ensure all threads are working equally. These tests show best performance with 4 MPI processes, above this it appears that the overhead of additional MPI processes causes performance to deteriorate. 
\FloatBarrier

\subsubsection{Multiple Node Scalability}
\label{sect:multinodescale}

\begin{figure} 
\centering
\fbox{
\includegraphics[height=5.0cm]{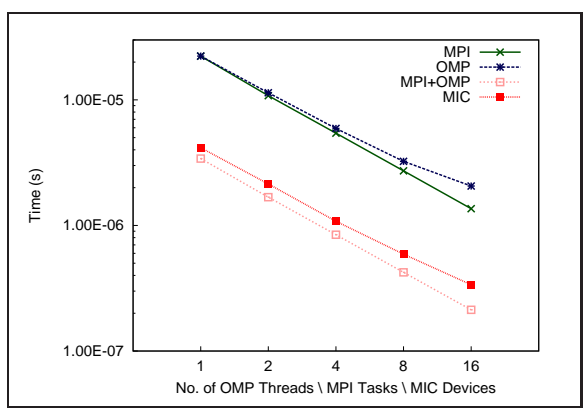}
}
\caption{Scalability: Per-Particle processing time for varying models from serial to highly parallel and Xeon Phi.}
\label{fig:scalability}
\end{figure}

We employ three sets of tests on the Xeon running from serial to highly parallel with multiple paradigms; a dual socket 16~core node exploited with OpenMP from a single to 16~threads, 1~to 16~single threaded MPI tasks with one task per core, and 1~to 16~ MPI tasks with one task per CPU and 8 OpenMP threads per task. These are compared with a final set with up to 16~Xeon Phi.

It can be seen that for the full Splotch code we currently achieve performance with one Xeon Phi roughly similar to one CPU parallelized with OpenMP. The use of a dataset larger than device memory causes a decrease in performance in comparison to that shown in Sect.~\ref{sect:singlenodeperf}. The non-linear scaling for the Xeon OpenMP implementation is due to locality issues during rendering, as threads access particles in memory according to their position when projected onto an image rather than their spatial locality. This is not an issue for the MPI implementation as each task renders particles independently of the other tasks, and there is no risk of non-local memory access. Scalability of the MIC is non-linear in the~8 to~16 range, this is due to the dataset not being large enough to fully exploit the power of the device when subdivided, we expect to see more linear scaling using larger datasets with device counts ranging above~8. 
\FloatBarrier

\subsection{Splotch: MIC vs. GPU}
\label{sect:micvsgpu}

Our experience of implementing the Splotch code for both GPUs and Xeon Phi allows to make a comparison of the performance we have achieved through similar expenditures of time and effort. In the case of the GPU we implement our ray casting algorithm in CUDA~\cite{Splotch03}, whilst in the case of the Xeon Phi we retain the original parallel model with OpenMP, and in both cases MPI can also be exploited. We use the same full dataset and host processors for performance tests. We render a single image of $1024^2$ pixels for six different camera positions, starting from very far and reaching progressively very close to the center of the simulation. In order to make a comparison we measure on a per-particle basis the total compute time (i.e. full algorithm minus data read and image output), the rendering kernel time, and the rasterization kernel time. 

\begin{figure} 
\centering
\fbox{
\includegraphics[height=5.0cm]{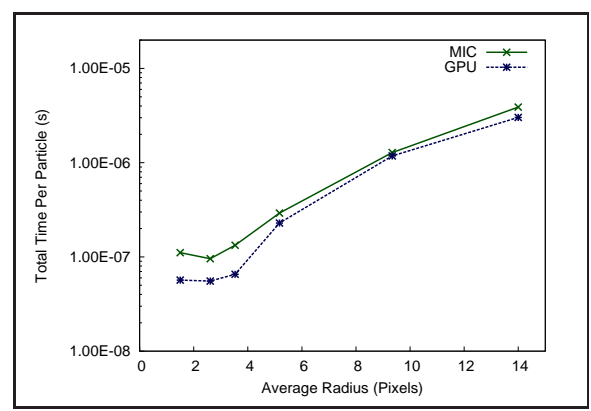}
}
\caption{Xeon Phi vs. NVIDIA K20X: comparison of the total compute times on a per-particle basis, using our OpenMP implementation for Xeon Phi vs our CUDA implementation for GPU.}
\label{fig:micgputotal}
\end{figure}

\begin{figure} 
\centering
\fbox{
\includegraphics[height=5.0cm]{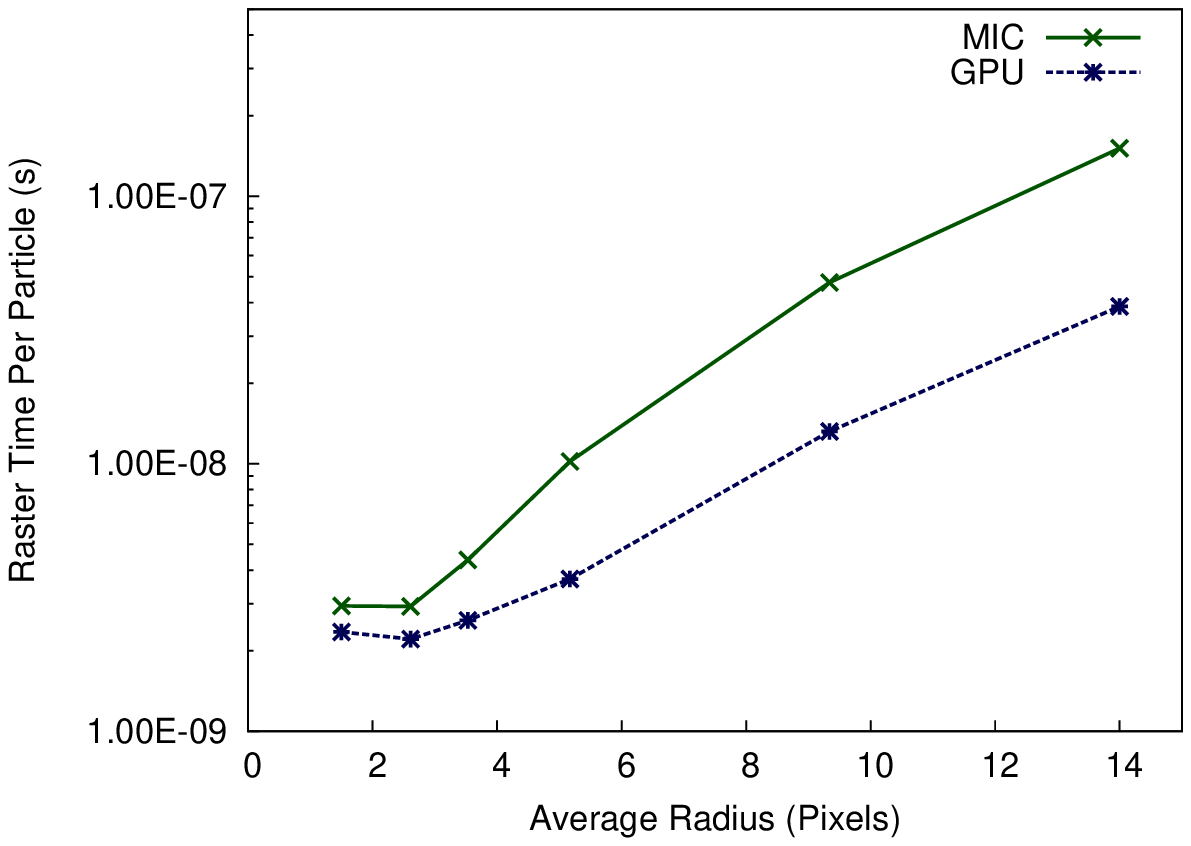}
}
\caption{Xeon Phi vs. NVIDIA K20X: comparison of the rasterization per-particle times, using our OpenMP implementation for Xeon Phi vs our CUDA implementation for GPU.}
\label{fig:micgpuraster}
\end{figure}

\begin{figure} 
\centering
\fbox{
\includegraphics[height=5.0cm]{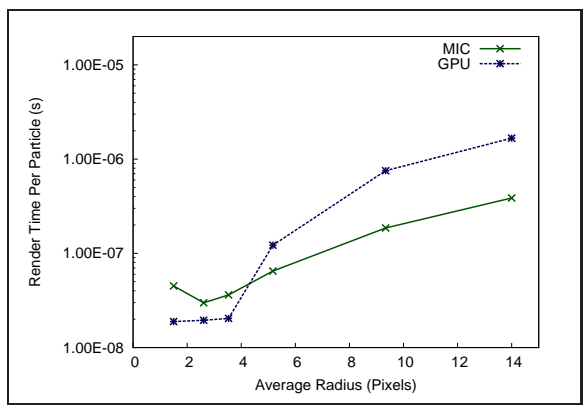}
}
\caption{Xeon Phi vs. NVIDIA K20X: comparison of the rendering per-particle times, using our OpenMP implementation for Xeon Phi vs our CUDA implementation for GPU.}
\label{fig:micgpurender}
\end{figure}

Figure~\ref{fig:micgputotal} shows total performance of a single NVIDIA K20X card versus a Xeon Phi, as  function of the average particle radius. The average radius, while not being the only factor affecting rendering time (see~\cite{Splotch03} for more detail), is a useful metric for comparison; a larger radius means the particle will affect more of the image and time to render will increase. It is clear that although the GPU implementation outperforms the Xeon Phi in all tests, for the larger radii the results are very close.

Figures~\ref{fig:micgpuraster} and \ref{fig:micgpurender} show kernel specific performance on a per particle basis. The performance difference shown in Fig.~\ref{fig:micgpuraster} is mostly attributed to the combination of the colorize and transform/filter kernels on the GPU. This means that for a large dataset where a significant portion of particles are inactive (i.e. off screen), as is the case for the tests with larger average radius, the kernel ends before considering these particles. To retain automatic vectorization of the transform kernel on the MIC the colorize kernel is run separately, and so all particles must be re-read and the active status tested, causing the colorize kernel to be dependent on the total number of particles as opposed to solely the number of active particles as is the case on the GPU. 

Figure~\ref{fig:micgpurender} shows performance comparison for the rendering phase. For a larger average radius, above two to three pixels, the MIC outperforms the GPU for rendering. This is due to the MIC algorithm being more suited to scenarios where a particle may affect a large portion of the image, as particles can be rendered by multiple threads when affecting multiple tiles. For the GPU, this is not possible as each tile is rendered by different CUDA blocks, therefore when a particle affects more than one tile it must be transferred back to the host and rendered (see~\cite{Splotch03} for more on this).  

The GPU performs very well in the lower radii range due to the fact that the large majority of particles are processed by the CUDA thread blocks and only a few of them are left to the CPU. Furthermore a specific one particle per thread approach is used with point-like particles which is ideal for the hardware. The MIC performance decreases in the case where a considerable portion of the image is unused (e.g. with a point of view far from the computational box center, as in Fig.~\ref{fig:visualization} \emph{left}). The current decomposition method does not effectively load balance this distribution of particles and requires improvements to account for such situations. 

\FloatBarrier

\subsection{Discussion}
\label{sect:discussion}
The results gathered so far demonstrate that in some areas of code the MIC architecture excels well beyond the host capabilities, although in others a fair amount of modification is necessary to gain acceptable performance levels, which is expected of a highly parallel architecture such as this. It is recommended to make extensive use of the optimization guides provided by Intel, and in order to achieve best performance rely not only on automatic vectorization but manual insertion of intrinsics also. Memory management is also key to performance; use of MPI based offload is shown to mitigate some overheads, similar to others' experience (e.g.~\cite{GadgetMicPrace}). Issues regarding scalable memory allocation, which may not be apparent with an identical implementation on a Xeon CPU, can be greatly improved by use of a thread-aware allocation mechanism as demonstrated in Sect.~\ref{sect:memusage}.

This work has focused purely on the offload model of exploiting the Xeon Phi, however the optimizations performed here can be effective not only to offload processing, but also to native processing. In observation of the future plans of Intel, in particular the second generation Xeon Phi product codenamed \emph{Knights Landing}~\cite{KnightsLanding}, we believe greater performance will be seen moving to a native model and utilizing the device as a processor in its own right. The improvements to the architecture in the second generation will remove many barriers to performance; the ability to function as a standalone processor with direct access to large memory will remove costly PCIe based data transfer, and the move to Atom based cores will allow for more advanced architectural features to be exploited, e.g. out of order execution, providing improved serial performance.

One reason for the lack of extensive profiling with Intel VTune throughout this work is that the authors experience of the current version is fairly positive for the command line, however the GUI was often unstable during remote offload profiling and so difficult to work with. It will be of interest to see if features provided in Intel Parallel Studio 15 such as revamped compiler optimization reports and support for the new OpenMP 4.0 specification provide further benefit to the code, e.g. with features such as the \textit{teams} construct, along with improvements to the level of profiling possible for Xeon Phi applications, e.g. the new Advanced Hotspots analysis method.  

The comparison of GPU and Xeon Phi shows that our GPU implementation currently outperforms the Xeon Phi in most areas. However Xeon Phi results are in some scenarios close or even better  than those of the GPU, as, for instance, when particle distributions with a large average radius are processed.The GPU architecture, in fact, is not suitable to render particles affecting large fractions of the image. Consequently, such particles have to be moved back to and processed by the CPU, with a strong performance penalty. This penalty is avoided by the Xeon Phi, whose  computing capabilities can be exploited for rendering particles of any size. It should be noted that the main barrier to performance for Xeon Phi is memory allocation. In each of the tests for MIC, memory allocation makes up at least half of overall time, and in the case where a Xeon Phi could allocate memory as fast as a Xeon the total performance outperforms the GPU in most tests.

\section{Conclusions}
\label{sect:conclusions}

In this paper we document our process of porting and optimizing Splotch, a visualization algorithm targeting large scale astrophysical data, to the Xeon Phi coprocessor. We explain the background of Splotch, and the efforts to achieve performance in an offloading model on the Xeon Phi by modifying our original OpenMP and MPI parallelized code to exploit Intel's LEO. We discuss in detail the optimizations performed through use of environment variables, thread affinity, vectorization (automatic and manual), and memory management. We profile offloaded code using a small wrapper around the Performance API for directly measuring hardware events, and share our experiences throughout the porting and optimizing process.

We run tests with multiple Xeon Phi vs a dual socket Xeon 2670, and record strong performance gains in individual kernels, i.e. 9x dual socket performance for the transform and color kernels. The algorithm as a whole proves comparable to the dual socket Xeon for datasets that fit within device memory, and a single socket for larger datasets. We contrast our achieved performance against that of our CUDA implementation on NVIDIA K20X graphics processors, and find that our GPU implementation currently outperforms the MIC implementation, albeit only marginally in some areas, with the MIC being particularly suited to the rendering stage. 

We also compare the experience of porting Splotch to Xeon Phi and GPUs. The MIC architecture, more ``traditional'' than that of the GPU, does not require extensive changes in the core algorithms, hence the design phase is simplified. Furthermore, the refactoring for the Xeon Phi is made easier by the ability to program with regular tools and familiar paradigms (OpenMP, MPI etc). On the other hand, performance tuning is, in most cases, highly demanding. Overall, the resulting time and effort needed to enable Splotch to run efficiently on a Xeon Phi is comparable to that needed for the GPU. With the introduction of coming models of Xeon Phi, i.e. Knights Landing, this may change in the near future. However, it is likely developments in GPU architecture will also start to reduce critical barriers to performance (e.g. NVLink), and further developments may simplify code development for heterogenous systems exploiting both architectures (e.g. OpenACC), we believe both versions can also benefit from further development to reach optimal performance. 

The Splotch code is now ready to effectively exploit new supercomputing systems making use of Xeon Phi devices. We intend to continue maintaining and developing the code and updating to exploit new hardware features when possible to continue to allow Splotch to utilize heterogenous systems in the best possible manner.

\begin{acks}
We would like to acknowledge CSCS-ETHZ, Lugano, Switzerland, for hosting author Timothy Dykes as part of their internship program, Klaus Dolag (University Observatory Munich) and Martin Reinecke (MPA Garching) for discussions and data for performance analysis. 
%Anonymized
\end{acks}

\begin{funding}
This research received no specific grant from any funding agency in the public, commercial, or not-for-profit sectors.
\end{funding}
    
%\begin{thebibliography}{99}
\bibliographystyle{sagev}
\bibliography{SplotchXeonPhi}

%\end{thebibliography}

\end{document}